\newif\ifnew\newtrue
\ifnew
     \documentclass[12pt]{article}
\usepackage[dvips]{epsfig}
\usepackage{latexsym}

  \else
     \documentstyle[12pt]{article}
     \newcommand{\imath}{\i}
     \newcommand{\jmath}{\j}
\fi

\csname @addtoreset\endcsname{equation}{section}

\newcommand{\eqn}[1]{(\ref{#1})}

\newsavebox{\uuunit}
\sbox{\uuunit}
    {\setlength{\unitlength}{0.825em}
     \begin{picture}(0.6,0.7)
        \thinlines
        \put(0,0){\line(1,0){0.5}}
        \put(0.15,0){\line(0,1){0.7}}
        \put(0.35,0){\line(0,1){0.8}}
       \multiput(0.3,0.8)(-0.04,-0.02){12}{\rule{0.5pt}{0.5pt}}
     \end {picture}}

\newcommand{\be}{\begin{equation}}
\newcommand{\ee}{\end{equation}}
\newcommand{\bea}{\begin{eqnarray}}
\newcommand{\eea}{\end{eqnarray}}
\newcommand{\nn}{\nonumber \\}
\newcommand{\ba}{\begin{array}}
\newcommand{\ea}{\end{array}}
\newcommand{\NP}[1]{Nucl.\ Phys.\ {\bf #1}}

\newcommand{\PRL}[1]{Phys.\ Rev.\ Lett.\ {\bf #1}}

\newcommand{\bfe}{\mbox{\boldmath $E$}}
\newcommand{\bfb}{\mbox{\boldmath $B$}}
\newcommand{\bfna}{\mbox{\boldmath $\nabla$}}
\newcommand{\bfd}{\mbox{\boldmath $D$}}
\newcommand{\bfeh}{\mbox{\boldmath $\hat{E}$}}
\newcommand{\bfbh}{\mbox{\boldmath $\hat{B}$}}
\newcommand{\bfah}{\mbox{\boldmath $\hat{A}$}}

\newcommand{\ah}{\mbox{$\hat{A}$}}
\newcommand{\fh}{\mbox{$\hat{F}$}}
\newcommand{\xh}{\mbox{$\hat{X}$}}
\newcommand{\yh}{\mbox{$\hat{Y}$}}
\newcommand{\ze}{\mbox{$\theta$-evolution}}
\newcommand{\nc}{non-commutative}
\newcommand{\wv}{worldvolume}
\newcommand{\bg}{background}
\newcommand{\sw}{Seiberg-Witten}

\newcommand{\z}{\mbox{$\theta$}}
\newcommand{\sz}{\mbox{$SL(2,\bZ)$}}
\newcommand{\bfz}{\mbox{\boldmath $\theta$}}
\newcommand{\bfr}{\mbox{\boldmath $\rho$}}
\newcommand{\bfxi}{\mbox{\boldmath $\xi$}}
\newcommand{\bfs}{\mbox{\boldmath $\sigma$}}
\newcommand{\bft}{\mbox{\boldmath $t$}}
\newcommand{\bfss}{\mbox{\boldmath $s$}}

\newcommand{\fc}{\frac}
\newcommand{\oh}{\frac{1}{2}}

\newcommand{\ap}{\mbox{$\alpha'$}}
\newcommand{\pq}{\mbox{$(p,q)$}}
\font\mybb=msbm12 at 12pt
\def\bb#1{\hbox{\mybb#1}}
\newcommand{\bZ}{\bb{Z}}

\def\lh {\hat{\lambda}}

\def\a {\alpha}
\def\b {\beta}

\def\d {\delta}
\def\sac{\, , \qquad}
\def\pa {\partial}

\begin{document}


\begin{titlepage}
\begin{flushright}
UB-ECM-PF-00/01 \\
hep-th/0002020
\end{flushright}
\vspace{.5cm}
\begin{center}
\baselineskip=25pt {\LARGE\bf Non-commutative vs. Commutative \\
Descriptions of D-brane BIons} \vskip 0.3cm \vskip 10.mm {\Large
David Mateos $^{*}$} \\ \vskip 2cm {\it

  Departament ECM, Facultat de F\'{\i}sica,\\
  Universitat de Barcelona and Institut de F\'{\i}sica d'Altes Energies,\\
  Diagonal 647, E-08028 Barcelona, Spain
\\ \vspace{6pt}
}
\end{center}
\vskip 5cm
\par
\begin{center}
{\bf ABSTRACT}
\end{center}
\begin{quote}
The $U(1)$ gauge theory on a D3-brane with non-commutative
worldvolume is shown to admit BIon-like solutions which saturate a
BPS bound on the energy. The mapping of these solutions to
ordinary fields is found exactly, namely non-perturbatively in the
non-commutativity parameters. The result is precisely an ordinary
supersymmetric BIon in the presence of a background $B$-field. We
argue that the result provides evidence in favour of the exact
equivalence of the non-commutative and the ordinary descriptions
of D-branes. 
\vskip 1.cm
\hrule width 5.cm 
\vskip 2.mm 
{\small \noindent $^*$ E-mail: mateos@ecm.ub.es}
\end{quote}
\end{titlepage}


\section{Introduction and Conclusions}


It has been recently appreciated \cite{Connes,DH,SW} that the dynamics
of D-branes in a constant background $B$-field
admits two equivalent descriptions: either in terms of an `ordinary'
gauge theory, or in terms of a gauge theory on a
{\it non-commutative} worldvolume. The two descriptions arise as a
result of using two different regularizations for the open string
worldsheet theory \cite{SW}. To provide evidence that both
descriptions are equivalent {\it non-perturbatively} in the
non-commutativity parameters is the purpose of this paper.

If Pauli-Villars regularization is used, the effective action for the
massless open string fields enjoys an ordinary
$U(N)$ gauge symmetry, where $N$ is the number of overlapping branes.
This gauge symmetry is therefore commutative in the case that one single
D-brane is present, which will be the only case considered in the
present paper. The (bosonic) massless fields are a gauge
potential $A_\mu$ ($\mu=0,\ldots,p$) and a number of scalars
$X^i$ ($i=p+1,\ldots,9$). With Pauli-Villars regularization,
the parameters entering the effective
action are the closed string metric $g$, the closed string coupling constant
$g_s$ and the constant background $B$-field itself. Furthermore, the
whole dependence of the effective action on $B$ is accounted for by
writing it in terms of the modified field strength
${\cal F} \equiv F + B^\star$, where $F=dA$ and the superscript
`$^\star$' denotes the
pull-back to the brane \wv. This is the combination which
is invariant under the two $U(1)$ gauge symmetries which the open
string $\sigma$-model enjoys at the classical level: a first one under
which 
\be
\delta B = 0 \sac \delta A = d \lambda \,,
\label{l-symmetry}
\ee
and a second one under which 
\be
\delta B = d \Lambda \sac \delta A = - \Lambda^\star \,.
\label{L-symmetry}
\ee

On the contrary, if point-splitting regularization is used, the gauge
symmetry group of the effective action becomes non-commutative even in the
case of one single D-brane. The theory is now naturally formulated as a
gauge theory on a non-commutative worldvolume \cite{SW},
namely a worldvolume whose
coordinates $\sigma^\mu$ do not commute but satisfy
\be
[\sigma^\mu,\sigma^\nu] \equiv \sigma^\mu * \sigma^\nu -
\sigma^\nu * \sigma^\mu = i \z^{\mu\nu}\,.
\ee
The antisymmetric matrix \z\ appearing above measures the
non-commutativity of the theory and defines the so-called `$*$-product' as
\be
(f*g)(\sigma) \equiv \left. e^{\frac{i}{2}\, \theta^{\mu\nu}\,
\frac{\pa}{\pa \xi^\mu}\, \frac{\pa}{\pa \chi^\nu}} \,
f(\sigma+\xi) \, g(\sigma+\chi)\right|_{\xi=\chi=0}\,.
\label{def.of.*}
\ee
The only parameters entering the effective action when point-splitting
regularization is used are \z, an open string metric $G$ and an open
string coupling constant $G_s$, whose relation with the closed string
parameters is
\footnote{We will set $2 \pi \ap = 1$.}
\be
\z = ( g + B )^{-1}_{(A)} \sac
G= g- B g^{-1} B \sac
G_s= g_s {\left( \fc{\det(g+B)}{\det g} \right)}^{1/2}\,,
\label{parameters}
\ee
where $(A)$ stands for the antisymmetric part of the matrix.
All products of fields in the effective action are now $*$-products,
and \z\ enters the action only through the definition of the $*$-product.
The \nc\ field strength is defined as
\bea
\fh_{\mu\nu} &=& \pa_{[\mu} \ah_{\nu]} - i [\ah_\mu,\ah_\nu] \sac \nn
{[} \ah_\mu , \ah_\nu {]} &=& \ah_\mu * \ah_\nu - \ah_\nu * \ah_\mu \,,
\eea
where $\ah_\mu$ is the \nc\ gauge field. With this regularization, the
gauge symmetry under which the effective action is invariant is
\be
\delta \ah = d\lh + i [\lh,\ah] \sac \delta \fh = i [\lh,\fh]\,.
\ee

As explained in \cite{SW}, since both the ordinary and the \nc\
descriptions arise from different regularizations of the same
worldsheet theory, there should be a field redefinition which maps
gauge orbits in one description into gauge orbits in the other. 
This requirement, plus locality to any finite order in \z,
enabled the authors in \cite{SW} to establish the following system
of differential equations: 
\footnote{The two equations \eqn{evolution} are simply the dimensional reduction of
that for a ten-dimensional gauge field $\ah_M$ $(M=0,\ldots,9)$,
which was given in \cite{SW}.} 
\bea 
	\delta \ah_\mu = \delta\theta^{\a\b} \, \frac{\pa \ah_\mu}{\pa \theta^{\a\b}} &=& -
\frac{1}{4} \, \delta\theta^{\a\b} \left\{ \ah_{\a} , \pa_\b
\ah_\mu + \fh_{\b\mu} \right\} \, ,\nn 
	\delta \xh^i = \delta\theta^{\a\b} \, \frac{\pa \xh^i}{\pa \theta^{\a\b}} &=& -
\frac{1}{4} \, \delta\theta^{\a\b} \left\{ \ah_{\a} , \pa_\b \xh^i
+ D_\b \xh^i \right\}\,, 
	\label{evolution} 
\eea 
where $\{f,g\} \equiv f*g+g*f$ and $D_\mu X \equiv \pa_\mu \xh - i
[\ah_\mu,\xh]$. These equations, which we will
call \ze\ equations, determine how the fields should change when
\z\ is varied, in order to describe equivalent physics. Their
integration provides the desired map between two descriptions with
different values of \z, to which we will refer as the Seiberg-Witten map.

The fact that two apparently so different descriptions can be
equivalent is certainly remarkable. The authors in \cite{SW}
provided a direct check that this is indeed the case by showing that
the effective action in one description is mapped to the effective
action in the other by the Seiberg-Witten map.
However, they worked in the approximation of slowly-varying fields, 
which consists of neglecting all terms of order $\pa F$ (or $\pa^2 X$).
This approximation was used at two different stages.
First, when the effective action for the massless fields
on the brane was taken to be the Dirac-Born-Infeld (DBI)
action. Indeed, this action can be derived from string theory
precisely by neglecting such terms. 
Second, this approximation was also used to simplify the \sw\ map considerably.

Although this check provides direct evidence in favour of the
equivalence of the two descriptions, it would be desirable to have an exact
proof. This would consist of three steps.
First, one would have to determine the effective action in each
description exactly, namely to all orders in \ap. Second, one would have to
integrate \eqn{evolution} also exactly, namely to all orders in
\z. Third, one would have to substitute the change of variables in one
action and see that the other one is recovered.
Of course, this procedure is impossible to put into
practice, but we will see that it is still possible to provide some 
evidence that the \sw\ map works non-perturbatively in \z.

The idea is as follows. If one exact effective action is mapped to
the other by the \sw\ map, then a classical solution of one
action should also be mapped to a classical solution of the other.
Of course, since we do not know the exact effective actions, we do
not know any non-trivial exact solutions either, except for one
case: the BIon \cite{CM,Gibbons,BPSbounds}. This is a
1/2-supersymmetric solution of the ordinary DBI theory of a
D-brane. In general, supersymmetric solutions of the worldvolume
theories of branes have a natural interpretation as intersections
of branes. The BIon is the prototype of this fact: it is
the worldvolume realization of a fundamental string ending on a
D-brane. What makes the BIon solution special is that, although
originally discovered as a solution of the DBI action
\cite{CM,Gibbons}, it has been shown to be a solution of the exact
effective action to all orders in \ap\ \cite{Thorlacius}. Perhaps
this should not be surprising since, after all, the fact that a
fundamental string can end on a D-brane is the defining property
of D-branes \cite{Polchinski}.

When no \bg\ $B$-field is present, the string ends orthogonally on
the brane (see figure 1(a)). When a constant {\it
electric}~\footnote{An electric 
$B$-field leads, through \eqn{parameters}, to an
electric \z\ in the \nc\ theory. There is some controversy about 
whether or not such a theory makes sense. 
We will postpone the discussion of this issue until the last section.} 
$B$-field is turned on, supersymmetry requires the string to
tilt a certain angle $\gamma$ determined by $B$ (see figure 1(b)). This
can be intuitively understood, because the background $B$-field
induces a constant electric field on the brane~\footnote{Fundamental 
strings ending on D-branes with constant
electric fields on their worldvolumes have also been studied in
\cite{Has-elec}. We will clarify the relationship of the results in
\cite{Has-elec} with ours at the end of section \ref{ordinary-BIons}. \label{Has}}.
Since the endpoint of the string is electrically charged, 
the string is now forced to tilt
in order for its tension to compensate the electric force on its
endpoint.

Our strategy will be to identify a BIon-like solution of the
effective action in the \nc\ description with the value of \z\
determined by $B$ as in \eqn{parameters}. Since the exact
effective action is not known, we will work with the lowest order
approximation in \ap\ (see \eqn{nc-action}). We will then integrate the \ze\ 
equations exactly to find what this \nc\ BIon is mapped to in the ordinary
description. The result is that it is mapped to a
tilted ordinary BIon, as described above, and that the tilting
angle $\gamma$ agrees precisely with the value determined by 
$B$~\footnote{The fact that tilted brane configurations are related to
monopoles and dyons in \nc\ gauge theories was first pointed out in
\cite{Has-Has} by working to first order in \z.}.

Our result can be interpreted in two ways. On one hand, if one accepts that the
equivalence between the ordinary and the \nc\ descriptions, as
determined by the \sw\ map, is valid non-perturbatively in \z,
then the result proves that the \nc\ BIon 
is a solution of the exact \nc\ effective action,
namely it is a solution to all orders in \ap.
On the other hand, motivated by the defining property of D-branes
mentioned above, one could directly conjecture that the \nc\ BIon is
a solution of the exact \nc\ effective action. In this case, one could
regard our result as evidence in favour of the exact equivalence
of both descriptions beyond the perturbative level in \z.
Whichever point of view one adopts, the result sheds some light on
another related question raised in \cite{SW}: the convergence of the
series in \z\ generated by the equations \eqn{evolution}. As far as we
are aware, this series has only been shown to converge for the case of
constant field strength \cite{SW}. Our example shows that it converges
in a less simple case.

In this paper we will concentrate on the case of D3-branes, but
the analysis for BIons applies to D-branes of 
arbitrary dimension.
The reason for this restriction is that we will briefly
comment on solutions more general than BIons, namely on dyons. These
are solutions of a D3-brane \wv\ theory carrying both electric and
magnetic charges which constitute the \wv\ realization of
$(p,q)$-strings ending on a D3-brane.
They fill out orbits of the \sz\ duality group of type IIB
string theory which, on the worldvolume of a D3-brane, becomes an
electromagnetic duality group. In the last section we will briefly comment
on the possible role of \sz\ in the \nc\ description of D3-branes. 
Section \ref{ordinary-BIons} is mainly a review. Sections
\ref{nc-dyons-section} and \ref{nc-vs-c} contain original results.

\begin{figure}
\begin{center}
	\begin{tabular}{c@{\hspace{2cm}}c}
		\epsfig{file=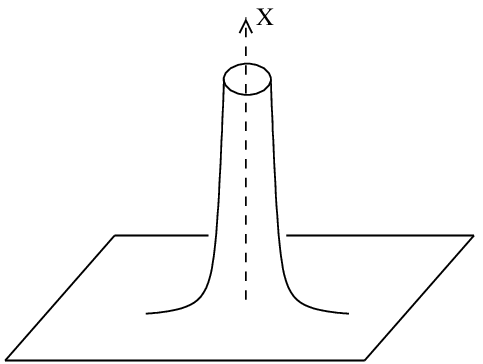, height=5cm, width=5cm} &
		\epsfig{file=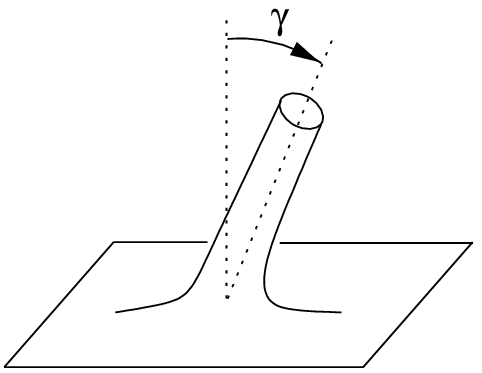, height=5cm, width=5cm} \\
		(a) & (b)
	\end{tabular}
\end{center}
\caption{The worldvolume realization of a fundamental string ending on
		a D-brane: (a) in the absence of $B$-field, and (b) in
		the presence of an electric $B$-field.}
\end{figure}


\section{Ordinary BIons in a background electric $B$-field}
\label{ordinary-BIons}

In this section we will consider the ordinary description of the
\wv\ theory of a D3-brane in the approximation of slowly-varying
fields. The action is therefore the DBI action. We will begin by
reviewing its dyonic solutions in the absence of a background
$B$-field. Then we will concentrate on the case of purely electric
solutions, the so-called `BIons'. Finally we will see
how the BIons are modified when a constant electric background
$B$-field is turned on. In this section we will work in the static gauge $X^\mu
\equiv \sigma^\mu$ ($\mu = 0, \ldots , 3$). $X^\mu$, together with
the scalar fields on the brane $X^i$, ($i=4,\ldots,9$), are
target-space Cartesian coordinates, that is, the closed string metric $g$ is
assumed to take the form $g_{MN} = \eta_{MN}$ ($M,N=0,\ldots 9$)
when expressed in these coordinates. Throughout this paper we will only
allow for one scalar field to be excited. Therefore, we will consider
the target-space to be effectively five-dimensional.

The dyonic solutions we wish to describe are the worldvolume
realizations of \pq -strings ending orthogonally on the D3-brane.
Since this is a spacetime-supersymmetric configuration, we have to
look for a worldvolume-supersymmetric solution. Supersymmetric
solutions of the D3-brane DBI action can be found either by
imposing directly preservation of supersymmetry
\cite{CM,Thorlacius.Peet} or by looking for BPS bounds on the
energy \cite{BPSbounds}. In either case, the BPS equations for
static 1/2-supersymmetric dyons carrying electric and magnetic
charges $p$ and $q$ respectively, are \cite{BPSbounds,CM}
\be
\bfe = \sin\a \, \bfna X \sac \bfb = \cos\a\, \bfna X \,,
\label{BPSdyons} 
\ee 
where \bfe\ and \bfb\  are the electric and
magnetic fields on the brane, $\tan \a = p/q$, and $X$ is the only
scalar field involved in the solution. The bound on the energy ${\cal
E}$ which the solutions of \eqn{BPSdyons} saturate is \cite{BPSbounds}
\be
{\cal E} \geq \sqrt{Z_{el}^2 + Z_{mag}^2}\,,
\ee
where the electric and magnetic charges above are
\be
Z_{el} = \int_\Sigma d^3\sigma \, \bfe \cdot \bfna X \sac 
Z_{mag} = \int_\Sigma d^3\sigma \, \bfb \cdot \bfna X \,, 
\label{com-charges} 
\ee 
and $\Sigma$ is the D3-brane worldspace.
Since both \bfe\ and \bfb\ are divergence free,
\be
\bfna \cdot \bfe = 0 \sac \bfna \cdot \bfb = 0 \,,
\label{com-id}
\ee
(the former as a consequence of the Gauss law and the latter
because of the Bianchi identity), these charges can be rewritten as
surface integrals over the boundary of the brane worldspace:
\be
Z_{el} = \int_{\pa \Sigma} d{\bf S} \cdot \, X \bfe \sac
Z_{mag} = \int_{\pa \Sigma} d{\bf S} \cdot \, X \bfb \,. 
\label{surface-charges}
\ee
This ensures that the charges are `topological', in the
sense that they only depend on the boundary conditions imposed on the fields.
It is this topological nature which guarantees that the saturation
of the bound automatically implies the equations of motion.

We see from \eqn{BPSdyons} and \eqn{com-id} that
$X$ must be harmonic, that is, $\nabla^2 X= 0$. Given a harmonic
function $X$, the electric and magnetic fields are determined by
(\ref{BPSdyons}). A dyon is then associated with an isolated
singularity of $X$.

We will concentrate for the rest of this section on the BIon. It corresponds to $\sin
\a = \pm 1$ in \eqn{BPSdyons}, and therefore satisfies 
\be
\bfe = - \bfna X \,,
\label{BPS-BIon}
\ee
where we have chosen the minus sign for convenience.
The most general $SO(3)$-symmetric solution is then given (up to a
gauge transformation) by
\be
X = A_0 = \frac{e}{4 \pi |\bfs|} \sac \mbox{\boldmath $A$} = 0\,,
\label{BIon}
\ee
where $\bfs=(\sigma^a)$, $a= 1,2,3$. It corresponds to a fundamental string ending
orthogonally on the brane at $\bfs=0$, as depicted in figure 1(a).
As mentioned above, \eqn{BIon} is a solution of (classical) string
theory to all orders in $\a'$ \cite{Thorlacius},
that is, when all corrections to the DBI action involving higher
derivatives of the fields are taken into account.

Since the BIon \eqn{BIon} saturates the BPS bound \eqn{BPS-BIon}, its energy
equals its charge. Furthermore, the latter is easily calculated with
the help of \eqn{surface-charges}. The boundary of the BIon worldspace consists of
a two-sphere at $|\bfs| \rightarrow \infty$ and another at
$|\bfs| \rightarrow 0$. The surface integral over the former
vanishes. Over the latter, it diverges. We can regularize it by
integrating over a small two-sphere $S_\epsilon$ of radius $\epsilon$. Since $X$ is
constant on this sphere we are left with 
\be
{\cal E}= |Z_{el}|= \lim_{\epsilon \rightarrow 0} \left| X(\epsilon)\,
\int_{S_\epsilon}  d{\bf S} \cdot \bfe \right| = 
e \, \lim_{\epsilon \rightarrow 0} X(\epsilon) \,.
\label{BIon-energy-X}
\ee
This shows that the energy of the BIon is precisely the energy of an
infinite string of constant tension \cite{CM,BPSbounds}. To compare with the BIon in
the presence of a $B$-field and with the \nc\ BIon, it will be
convenient for us to rewrite \eqn{BIon-energy-X} as
\be
{\cal E} = |Z_{el}| = z_{el} \, \lim_{\epsilon \rightarrow 0} I(\epsilon)\,,
\label{BIon-energy-I}
\ee
where
\be
z_{el} = e^2 \sac I(\epsilon) = \fc{1}{4 \pi \epsilon} \,.
\ee

Let us now see how the BIon solution is modified when a constant
electric background $B$-field is turned on. We  assume that the
target-space ten-vector $B_{0M}$ no longer vanishes (but we still
impose the restriction that $B$ has no magnetic components).
In general, any non-vanishing component of $B$ along directions `transverse' to the
brane can be gauged away. In our case, however, we have to be careful, because
we are looking for a configuration in which one scalar field is
excited. In other words, the worldspace of the brane is not a flat
three-plane extending along the directions 1, 2 and 3. Therefore, we need to
consider the component of $B$ along the direction labelled by $X$ (see
figure 2(a)), to which we will refer as $B_X$. Without loss of generality, we can take
the component of $B$ along the 123-space to point along the
1-direction. We will denote this component by $B_\sigma$. The remaining
components of $B_{0M}$ can be gauged away and we will
therefore set them to zero.
\begin{figure}
\begin{center}
	\begin{tabular}{c@{\hspace{2cm}}c}
		\epsfig{file=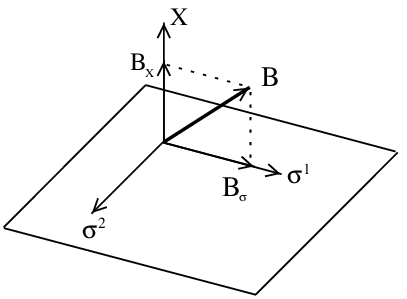, height=5cm, width=5cm} &
		\epsfig{file=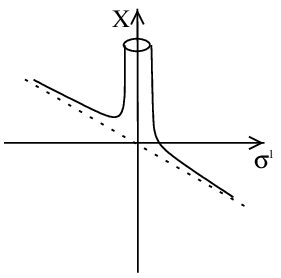, height=5cm, width=5cm} \\
		(a) & (b)
	\end{tabular}
\end{center}
\caption{(a) Decomposition of the electric $B$-field,
		and (b) projection on the $\sigma^1$-$X$ plane of the
		`tilted'  BIon \eqn{B-solution}.}
\end{figure}

In the presence of the $B$-field, the BIon BPS equation which
guarantees the preservation of some fraction of supersymmetry must
be modified: the field strength $F=dA$ is replaced by ${\cal
F}=F + B^\star$. This can be easily understood, since the
supersymmetry condition has to be gauge-invariant under the two
$U(1)$ gauge symmetries \eqn{l-symmetry} and \eqn{L-symmetry}.
Thus, \eqn{BPS-BIon} becomes
\be
{\cal F}_{0a} = - \pa_a X \,, 
\label{B-BPS}
\ee
whose solution is now
\be
A_0 = \frac{e}{4 \pi |\bfs|} \sac \mbox{\boldmath $A$} = 0 \sac 
X= \frac{1}{1+B_X} \, \fc{e}{4 \pi |\bfs|} - \fc{B_\sigma}{1+B_X} \, \sigma^1 \,. 
\label{B-solution} 
\ee 
Note the appearance of
a term linear in the worldspace coordinate $\sigma^1$ in the expression for
the scalar field. This is the term responsible for the tilt of the
string (see figure 2(b)). Indeed, although the spike coming out of the
brane at $|\bfs|=0$ still points along the $X$-axis and the brane
worldspace is still asymptotically flat, the latter no longer
asymptotically extends along the 1-direction.

The energy of the solution \eqn{B-solution} is computed analogously as
we did with \eqn{BIon}. The surface integral at infinity still
vanishes (the term in the scalar field which is linear in $\sigma^1$ does
not contribute because it changes sign under $\bfs \rightarrow -
\bfs$). Thus, we obtain again \eqn{BIon-energy-I}, but with $z_{el}$ now
given by
\be
z_{el} = \fc{e^2}{1 + B_X}
\label{z_el-in-B}
\ee

We would like to close this section with some clarifications. 
The first one is that it might appear that \eqn{B-solution} is not the
most general solution of \eqn{B-BPS}: we could have included terms
linear in the worldspace coordinate both in $X$ and in $A_0$, with
appropriate coefficients such that \eqn{B-BPS} be satisfied. However,
this apparently more general solution is related to \eqn{B-solution} by
a gauge transformation of the type \eqn{L-symmetry}; they are
therefore physically equivalent. (Admittedly, such a transformation
would shift the value of $B$, but this is allowed since $B$ is
generic in our analysis.) In particular, by choosing the target-space
one-form as
\be
\Lambda = \left( B_\sigma \sigma^1 + B_X X \right) \, d\sigma^0\,,
\ee
the solution \eqn{B-solution} in the presence of a non-vanishing
$B$-field is mapped to a configuration with $A_0=X$ (where $X$ is
still given by \eqn{B-solution}) and vanishing $B$-field. This configuration
satisfies \eqn{BPS-BIon}, and is precisely of the form considered in
\cite{Has-elec}. These considerations therefore clarify the
relation between the solution studied in \cite{Has-elec} and the one
we have presented: they are related by a gauge symmetry of the
theory; whether the constant electric field on the brane is induced
from the background or whether it arises from the worldvolume gauge field itself
is a matter of gauge choice. 
We have chosen to work in a gauge in which $B\neq 0$ since it is in
this case that a \nc\ alternative description exists.

The second remark is that the proof in \cite{Thorlacius} that the BIon
\eqn{BIon} is an exact solution to all orders in \ap\ assumed the
absence of a background $B$-field. Therefore, one might question
whether the result also holds for our BIon \eqn{B-solution}.
The answer is affirmative, because, as we have just explained,
\eqn{B-solution} is related by a gauge symmetry of the theory to the
configuration studied in \cite{Has-elec} in which $B=0$. As pointed
out in \cite{Has-elec}, this latter configuration is indeed a solution
to all orders in \ap, because it satisfies \eqn{BPS-BIon}, which 
was the only assumption in \cite{Thorlacius} 
(as opposed to any assumption concerning the
specific form of the solution, such as \eqn{BIon}).


\section{Non-commutative D3-brane Dyons}
\label{nc-dyons-section}


In this section we will first establish the non-commutative
version of the BPS equations \eqn{BPSdyons}. Then we will
construct an exact solution for the case of purely electric
charge, which we will refer to as the `\nc\ BIon'. In the next
section, we will show that this solution is mapped to the
ordinary BIon \eqn{B-solution} in a $B$-field by the \sw\ map.
Therefore, the metric we must use here is the open string
metric $G$ as determined by $g$ and $B$ in \eqn{parameters}. This is
important since, although both $g$ and $G$
are flat, they are not necessarily simultaneously diagonal in the
same coordinate system. In this section it will be convenient to
work in a system in which $G$ takes the simplest form
$G_{MN}=\eta_{MN}$ (note that we are referring to the target-space
coordinate system, and that therefore it includes the scalar
field on the brane). The system in which $G$ takes this form is
obtained as follows (see figure 3). 
\begin{figure}
\begin{center}
	\epsfig{file=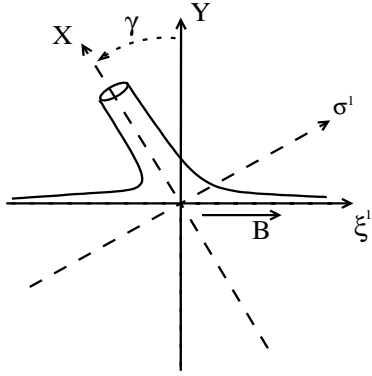, height=5cm, width=5cm} 
\end{center}
\caption{The convenient coordinate system in the analysis of the
		\nc\ BIon, and the projection to the $\xi^1$-$Y$ plane
		of the solution \eqn{xis-solution}.}
\end{figure}
Let us take the first worldspace
coordinate $\xi^1$ along the direction of $B$, and the scalar
field $Y$ to be orthogonal to it. In this system we have
\be
B=b \,d\xi^0 \wedge d\xi^1 \,, 
\label{Bb} 
\ee 
where $b$ is a positive constant, but $G$ does not yet take the desired
form. However, the simple rescaling
\be
\rho^0 = \sqrt{1-b^2}\,\xi^0 \sac \rho^1 = \sqrt{1-b^2}\,\xi^1
\label{rescaling} 
\ee 
brings $G$ into such a form: 
\be
ds^2_{(G)}= -d\rho_0^2 + d\rho_1^2 + d\rho_2^2 + d\rho_3^2 + dY^2
\ee 

In this coordinate system \z\ also takes a very simple form:
its only non-vanishing components are
\be
\z^{01}= - \z^{10} = b\,. 
\label{theta-rho} 
\ee 

After these preliminaries, we are ready to write down the \nc\ D3-brane \wv\
action. We will take it to be the lowest order approximation in \ap, namely
\be
S = \int d^{1+3}\rho \, \left( -\fc{1}{4} \hat{F}_{\mu\nu} *
\hat{F}^{\mu\nu} - \fc{1}{2} D_\mu \yh * D^\mu \yh \right)\,. 
\label{nc-action}
\ee
All indices above are contracted with $G$.

One might think that a Hamiltonian analysis of the \nc\ action is
required in order to obtain its energy functional, which we
certainly need to derive a BPS bound on the energy. This seems
difficult in view of the non-local nature of the action. 
(Note that this non-locality is not only in space but also in time,
since we will not impose the restriction $\z^{0a}=0$.) The problem
can be circumvented by noting that, although the presence of \z\
breaks Lorentz invariance, the theory is still translation-invariant. 
Therefore we can obtain the Lagrangian energy ${\cal E}$ as the
conserved quantity under time translations. The result is
\be
{\cal E} = \int_\Sigma d^3\rho \left( \fc{1}{2} \bfeh^2 + \fc{1}{2} \bfbh^2 +
\fc{1}{2} (D_0 \yh)^2 + \fc{1}{2} (\bfd \yh)^2 \right)\,,
\ee 
where \bfeh\ and \bfbh\ are the \nc\ electric and magnetic fields,
that is,
\be
\hat{E}_a = \fh_{0a} \sac \hat{B}_a = \oh \epsilon_{abc}\, \hat{F}_{bc}
\,.
\ee
For covariantly static configurations, namely with $D_0 \yh = 0$, the
energy can be rewritten as
\be
{\cal E} = \int_\Sigma d^3 \rho \left[ \oh \left( \bfeh - \sin \a \bfd \yh \right)^2
+ \oh \left( \bfbh - \cos \a \bfd \yh \right)^2 + \sin \a \bfeh \cdot
\bfd \yh + \cos \a \bfbh \cdot \bfd \yh \right]\,, 
\ee 
from which the desired BPS bound follows immediately:
\be
{\cal E} \geq \sqrt{Z_{el}^2 + Z_{mag}^2}\,. 
\ee 
The electric and magnetic charges above are the natural \nc\
generalizations of \eqn{com-charges}:
\be
Z_{el} = \int_\Sigma d^3\sigma\, \bfeh \cdot \bfd \yh \sac 
Z_{mag} = \int_\Sigma d^3\sigma\, \bfbh \cdot \bfd \yh \,. 
\label{nc-charges} 
\ee 
The bound is saturated precisely when the \nc\ BPS equations
\be
\bfeh = \sin \a \, \bfd \yh \sac \bfbh = \cos \a \, \bfd \yh \,
\label{nc-dyons} 
\ee 
hold~\footnote{The case with $\cos \a = 1$ in
\eqn{nc-dyons}, which corresponds to a
monopole, was studied in \cite{Brane-Config} in the context of a $U(2)$
gauge group. The \nc\ monopole equation was solved to first order in
\z. The solution exhibits a certain non-locality corresponding to the
tilt of the D-string ending on the brane.}. 
They are also the natural generalizations of their commutative
counterparts \eqn{BPSdyons}.

As well as the BPS equations, the Gauss law and the Bianchi
identity \eqn{com-id} should also be promoted to their \nc\
versions
\be
\bfd \cdot \bfeh = 0 \sac \bfd \cdot \bfbh = 0 \,. 
\label{nc-GB}
\ee 
A number of comments are in order here. The new
Bianchi identity follows immediately from the definition of $\fh$.
However, it is not clear to us whether it constitutes a locally
sufficient integrability condition (as it does in an ordinary
gauge theory) which ensures that $\fh$ can be written as the
covariant derivative of a gauge potential. The
Gauss law above is nothing else than one of the equations of
motion. However, a rigorous proof that it is a constraint in the
\nc\ theory would require a careful analysis, which is difficult
again due to its non-locality. Nevertheless, there are two
observations which are worth noticing. First of all, at least in
the case of purely magnetic \z, that is, when $\z^{0a} = 0$, the \nc\
theory becomes local in time and the Hamiltonian analysis is
straightforward \cite{Dayi}. In this case one can prove that the Gauss
law (which contains only first order time derivatives in its
Lagrangian form) becomes a Hamiltonian constraint.
Second, both the Bianchi identity and the Gauss law are required
in order to rewrite the electric and magnetic charges as surface
integrals:
\be
Z_{el} = \int_{\pa \Sigma} d{\bf S} \cdot \, \yh \bfeh \sac
Z_{mag} = \int_{\pa \Sigma} d{\bf S} \cdot \, \yh \bfbh \,. 
\ee
As in the ordinary case, this is what ensures that they have a
topological nature, which in turn guarantees that the saturation
of the bound automatically implies the equations of motion.

Now we are ready to finally write down the \nc\ BIon solution. It
is simple to check that the configuration
\be
\ah_0^{(\rho)} = \yh = \frac{q}{4 \pi |\bfr|} \sac
\bfah^{(\rho)}=0\,, 
\label{nc-BIon} 
\ee 
solves both the \nc\ BPS equations \eqn{nc-dyons} (with $\sin \a= -1$
as before) and
the \nc\ Gauss law \eqn{nc-GB}. In \eqn{nc-BIon} we have $\bfr =
(\rho^a)$, 
and the superscript
$(\rho)$ denotes that the above are the components of the
gauge potential one-form in the $\rho$-coordinate system, namely
that $\ah = \ah_0^{(\rho)} \, d \rho^0$. Its components in the
$\xi$-coordinate system are, by virtue of \eqn{rescaling},
\be
\ah_0^{(\xi)} = \sqrt{1-b^2} \, \ah_0^{(\rho)} \sac \bfah^{(\xi)}=0\,.
\label{A-rescaling} 
\ee
Hereafter we will denote $\ah_0^{(\rho)}$ simply by $\ah_0$ until the moment
in which we have to compare with \eqn{B-solution}. 

The solution \eqn{nc-BIon} is the announced \nc\ BIon. Its charge is
again easily computed as we did in the previous section. The result is
again \eqn{BIon-energy-I} with 
\be
z_{el} = q^2 \,.
\label{z_el-q}
\ee
and we see from \eqn{z_el-in-B} and \eqn{z_el-q} that we must choose
\be
q = \fc{e}{\sqrt{1+B_X}}
\label{q-vs-e}
\ee
in order for the charges (or equivalently, for the energies) 
of the \nc\ and of the ordinary BIon to agree.


\section{From Non-commutative to Commutative BIons}
\label{nc-vs-c}


Our goal in this section is to show that the solution
\eqn{nc-BIon} of the \nc\ theory (with non-commutativity parameter
\z\ determined by $B$ as in \eqn{theta-rho}) is precisely mapped
to the ordinary BIon \eqn{B-solution} in the presence of the $B$-field. We
should therefore integrate the \ze\ equations \eqn{evolution}
exactly, with the initial conditions \eqn{nc-BIon} for the fields
at the initial value of \z\ given in \eqn{theta-rho}. However,
there is an {\it a priori} obstacle for doing so which is worth
discussing. Indeed, the \ze\ equations constitute a system of
coupled partial differential equations, whose (local)
integrability conditions are that the crossed derivatives be
equal, namely that
\be
\frac{\pa^2 \ah_\mu}{\pa \z^{\gamma\rho} \pa \z^{\a\b}} =
\frac{\pa^2 \ah_\mu}{\pa \z^{\a\b} \pa \z^{\gamma\rho}} \, ,
\label{integrab} 
\ee 
and similarly for the scalar fields. As
proved in \cite{Asakawa}
\footnote{I thank Joan Sim\'on for drawing my attention to this
reference.}, 
these conditions are in general not
satisfied. This means that the evolution of the fields in
`\z-space' determined by integrating the equations \eqn{evolution}
between some initial and final values $\z_0$ and $\z_1$ depends on
the path followed from $\z_0$ to $\z_1$. The choice of path should
be made according to some physical input~\footnote{As explained in
\cite{Asakawa} there is no path dependence for the case of a $U(1)$
gauge group if terms of ${\cal O} (\pa \fh)$ are neglected.}.

In our case, we are interested in the \ze\ of the BIon
\eqn{nc-BIon} starting from an initial \z\ which is purely
electric, and ending at $\z=0$. This suggests that we should restrict
ourselves to the hyperplane $\z^{ab} = 0$ in \z-space, and 
consider paths contained within this hyperplane. This restriction can
be physically further motivated as follows. Suppose that we start
with a D3-brane in the absence of any $B$-field. In this situation
only the ordinary description is available. Now we smoothly turn
on the electric components of $B$. As soon as $B$ is different
from zero, both the ordinary and the \nc\ descriptions are
available. It follows from its definition \eqn{parameters} that,
in the latter description, \z\ will also be purely electric. We
can now imagine following the evolution of the fields as a
function of \z\ (or, equivalently, of $B$) as it increases. We see
that in this physical situation we are following a path which lies
in the $\z^{ab} = 0$ hyperplane.

The restriction $\z^{ab} = 0$ simplifies the \ze\ equations
dramatically, because it implies that the $*$-product of any two
time-independent functions $f$ and $g$ reduces to their ordinary
product, namely that $f*g = f g$. With this simplification the
equations \eqn{evolution} become 
\bea
\label{first} 
\fc{\pa \ah_0}{\pa \z^{b}} &=& - \ah_0 \, \pa_b
\ah_0 \, \\ 
\fc{\pa \ah_a}{\pa \z^{b}} &=& - \ah_0 \pa_b \ah_a +
\fc{1}{2} \ah_0 \pa_a \ah_b - \fc{1}{2} \ah_b \pa_a \ah_0 \,,
\label{second} \\ 
\fc{\pa \yh}{\pa \z^{b}} &=& - \ah_0 \, \pa_b
\yh \,, 
\label{third}  
\eea
where $\z^a \equiv \z^{0a}$. Note that in the above equations
$\pa_b \equiv \pa / \pa \rho^b$.

The unique solution of \eqn{second} with the initial condition
that $\ah_a$ vanishes at some initial value of \z\ is that it
vanishes for all values of \z, regardless of the chosen path.
Furthermore, taking the derivative of \eqn{first} and using
\eqn{first} itself, one can easily check that the crossed
derivatives of $\ah_0$ coincide. Finally, \eqn{first} and
\eqn{third}, together with the initial condition $\ah_0 = \yh$,
show that $\ah_0$ and $\yh$ remain equal for all values of \z. We
thus conclude that, in the hyperplane $\z^{ab} = 0$, the \ze\ of the
BIon \eqn{nc-BIon} is path-independent, and that to determine it
we need only solve \eqn{first} with the initial condition
\be
{\left. {\ah_0 (\bfr,\bfz)} \right|}_{\bfz=\bfz_0} = \fc{q}{4 \pi
|\bfr|} \,, \label{initial-con} 
\ee 
where $\bfz=(\z^a)$ and $\bfz_0=(b,0,0)$.

\eqn{first} constitutes a system of three quasi-linear
partial differential equations which can be solved by the method
of characteristics:\footnote{I am grateful to Emili Elizalde for help
at this point.} instead of solving \eqn{first} directly, one
introduces a new three-vector $\bft=(t^a)$ and solves
\be
\fc{\pa\rho^b}{\pa t^a}=\d^b_a \, \ah_0 \sac \fc{\pa\z^b}{\pa
t^a}=\d^b_a \sac \fc{\pa\ah_0}{\pa t^a}=0 \label{char-t} \ee with
initial conditions that depend on a further three-vector
$\bfss=(s^a)$, namely
\be
\bfr = \bfss \sac \bfz=\bfz_0 \sac \ah_0 = \fc{q}{4 \pi |\bfss|}
\qquad \qquad \mbox{at} \; \bft=0 \,. 
\label{char-s} 
\ee 
The physical interpretation of both \bft\ and \bfss\ will become clear
shortly.

\eqn{char-t} and \eqn{char-s} determine \bfr, \bfz\ and $\ah_0$ as
functions of \bft\ and \bfss. It is easy to see that the solution
is
\be
\bfr(\bft,\bfss)= \fc{q}{4 \pi |\bfss|} \, \bft + \bfss \sac \bfz
(\bft,\bfss) = \bft + \bfz_0 \sac \ah_0 (\bft,\bfss) = \fc{q}{4
\pi  |\bfss|} \,. \label{ts-solution} \ee The (local) inversion of
the first two relations above would determine $\bft(\bfr,\bfz)$
and $\bfss(\bfr,\bfz)$, which could be substituted into the third
one to obtain the desired solution $\ah_0(\bfr,\bfz)$. To see that
the function $\ah_0$ so determined satisfies \eqn{first} we
simply have to apply the chain rule and use \eqn{char-t} to
obtain:
\be
0 = \fc{\pa \ah_0}{\pa t^a} = \fc{\pa \ah_0}{\pa \z^b} \, \fc{\pa
\z^b}{\pa t^a} + \fc{\pa \ah_0}{\pa \rho^b} \, \fc{\pa \rho^b}{\pa
t^a} = \fc{\pa \ah_0}{\pa \z^a} + \ah_0 \, \fc{\pa \ah_0}{\pa
\rho^a} 
\ee 

It is also obvious from \eqn{char-s} that $\ah_0$
satisfies the required initial condition \eqn{initial-con}.

We see from \eqn{ts-solution} how to interpret \bft:
it is simply the difference between the initial and the final
values of the non-commutativity parameter. Therefore, in our case
we will be interested in examining the values of the fields at
\be
\bft = (-b,0,0) \,. 
\label{t-value} 
\ee 
\eqn{ts-solution} also shows that $\bfss=(s^a)$ are nothing else
than intrinsic coordinates on the D3-brane worldspace. Indeed,
recall that we argued that the scalar \yh\ and the potential
$\ah_0$ coincide for all values of \z. Therefore, for fixed \bft,
\eqn{ts-solution} determines the (static) embedding of the brane
in spacetime as
\be
\bfss \mapsto \left( \bfr(\bfss), \yh(\bfss) \right) 
\ee 
So we see that \eqn{ts-solution} provides the solution (namely the
gauge potential and the embedding of the brane in target-space) in
parametric form.

In order to check that the solution \eqn{ts-solution} at $\z=0$
is, as we claim, precisely the same as \eqn{B-solution}, we have to
express \eqn{ts-solution} in the same coordinate system as
\eqn{B-solution}.

First, we have to undo the rescaling \eqn{rescaling}, because the
closed string metric $g$ takes the Minkowski form in the
$\xi$-coordinates, but not in the $\rho$-coordinates. We thus
have, using \eqn{rescaling}, \eqn{A-rescaling}, 
\eqn{ts-solution} and \eqn{t-value}, that, at $\z=0$: 
\bea 
\xi^1 &=& -\fc{b}{\sqrt{1-b^2}} \, \fc{q}{4 \pi |\bfss|} +
\fc{1}{\sqrt{1-b^2}}\, s^1 \sac Y = \fc{q}{4 \pi |\bfss|} \,,
\label{xis-solution} \\
A_0 &=& \sqrt{1-b^2} \fc{q}{4 \pi|\bfss|} \sac 
\mbox{\boldmath $A$} = 0 \,.
\eea 
Note that we have dropped the `hats' on the fields, 
because the above are already their values
in the ordinary description. Note also that, although we did not write
the superscript ($\xi$) explicitly, $A_0$ and {\boldmath $A$} now denote the
components of the gauge potential in the $\xi$-coordinate system.

We have drawn the solution \eqn{xis-solution} (which is still
expressed in parametric form) in figure 3. For $|\bfss| \rightarrow
\infty$ we have $Y \sim 0$. Therefore the brane is asymptotically
flat and extends along the directions labeled by $\xi^1$, $\xi^2$
and $\xi^3$, as shown in the figure. On the contrary, for $|\bfss|
\rightarrow 0$, we see that
\be
Y \sim -\fc{\sqrt{1-b^2}}{b} \, \xi^1 
\ee 
This means that the
spike comes out of the region $|\bfxi| \rightarrow 0$ in a direction
at an angle $\gamma$ with the $Y$-axis (see figure 3 again), where
\be
\tan \gamma = \fc{b}{\sqrt{1-b^2}} 
\ee 
We have labelled this
direction by $X$, and the orthogonal direction in the $\xi^1$-$Y$
plane by $\sigma^1$. Recall that the ordinary BIon solution
\eqn{B-solution} is written in the static gauge and in a
coordinate system in which the spike points along the transverse
scalar field. In our case, this happens precisely in the $\sigma$-$X$
coordinate system. Therefore, the last step we have to take is to
rotate the solution \eqn{xis-solution} by an angle $\gamma$.
Defining
\be
\left(
  \begin{array}{c}
    \sigma^1 \\
    X
  \end{array} \right)
=
\left(  \begin{array}{cc}
    \cos\gamma & \sin\gamma \\
    -\sin\gamma & \cos\gamma
  \end{array} \right)
 \left( \begin{array}{c}
    \xi^1 \\
    Y
  \end{array} \right)
\label{rotation} 
\ee we find that $\sigma^1 = s^1$ and that 
\be
X= \fc{1}{\sqrt{1-b^2}}\, \fc{q}{4\pi |\bfss|} -
\fc{b}{\sqrt{1-b^2}} \, s^1 \,. 
\ee 
Finally, we see from \eqn{Bb} and \eqn{rotation} that
\be
B_\sigma = b \sqrt{1-b^2} \sac B_X= -b^2 \,.
\ee 
Therefore, using \eqn{q-vs-e}, we arrive at the final form of our
solution,  expressed in the static gauge:
\be
A_0 = \fc{e}{4 \pi |\bfs|} \sac X= \frac{1}{1+B_X} \fc{e}{4 \pi
|\bfs|} - \fc{B_\sigma}{1+B_X} \sigma_1 \,.
\ee 
It coincides precisely with \eqn{B-solution}, as we claimed.


\section{Discussion}


In this last section we wish to address a number of issues which
were not fully discussed in the text.

The first one concerns the interpretation of the scalar fields in
the \nc\ theory. In the ordinary description of a single D-brane,
the scalars unambigously determine the embedding of the brane in spacetime.
This is the reason why many phenomena in field theories
acquire a clearer geometrical interpretation when such
theories are realized as \wv\ theories of branes. In the \nc\
description, this interpretation is much less clear. The reason is
that the scalar fields, even in the case of one single D-brane,
are no longer gauge-invariant but gauge-covariant quantities;
namely they transform as $\delta \xh = i [ \hat{\lambda}, \xh]$.
This problem is related to the fact that all
gauge-invariant quantities in \nc\ gauge theories seem to be
non-local, obtained after integrating some gauge-covariant
quantity. Presumably, one can determine global properties of the
brane embedding from the \nc\ scalar fields, such as winding
numbers, etc., by integrating appropriate expressions, but not the
local details of the embedding. The reason why we did not have to
resolve this problem in our discussion of the \nc\ BIon is that we
were only interested in identifying a solution of the \nc\ theory
which was exactly mapped to the ordinary BIon in the presence of a
$B$-field. Since the Seiberg-Witten map maps gauge orbits into
gauge orbits and the ordinary scalar fields are gauge-invariant,
any scalar field configuration in the \nc\ theory which is
gauge-equivalent to \eqn{nc-BIon} would have also been mapped to
the ordinary BIon. We simply chose the simplest representative of
$\yh$ in its gauge-equivalence class.

The second point we wish to discuss here is whether it makes sense
to consider a non-commutativity matrix with non-vanishing electric
components
\footnote{I would like to thank Michael B. Green and Paul K. Townsend
for a conversation on this point.}.
In the ordinary description, it certainly makes sense to consider $B_{0a} \neq 0$,
which leads through \eqn{parameters} to $\z^{0a} \neq 0$.
Moreover, even if in one coordinate system $B$ has no electric
components, we can always choose to describe the physics in a
frame in which $B$ does have electric components. The only
question is whether the equivalence between the ordinary and the
\nc\ descriptions still holds {\it in this frame}. It seems 
that our result can be regarded as evidence in favour of the
affirmative answer to this question.

We would like to close this paper with a little digression on the role
of S-duality in the non-commutative theory, and more generally of
the full \sz\ duality group of type IIB string theory. Consider
the BIon solution in the ordinary description of D3-branes and in
the absence of any $B$-field. As we have explained, this is the
\wv\ realization of the spacetime configuration in which a
fundamental string ends on the D3-brane. The 
S-duality of string theory maps this configuration into one in which a D-string
ends on the D3-brane. Thus, from the \wv\ point of view of the
latter, S-duality is an inherited symmetry which corresponds to
electromagnetic duality. It maps the BIon into the monopole. These
two objects are therefore equivalent, in the sense that they are
related by a symmetry of the theory. One might think that this is
also the case in the \nc\ \wv\ description, perhaps with a further
exchange of the electric and the magnetic components of \z.
However, this is not true. The reason is that a BIon in the \nc\
theory corresponds, as we have seen, to a fundamental string
ending on the D3-brane in the presence of a $B$-field. S-duality
maps this configuration into a D-string ending on the D3-brane in
the presence now of a {\it Ramond-Ramond} $C$-field. Clearly, this
does not correspond to a monopole in the \nc\ theory, which should
instead correspond to a D-string ending on the D3 in the presence
of a $B$-field. These considerations raise the following interesting question.
There exists an \sz-covariant \wv\ action for
D3-branes coupled to a supergravity background \cite{Cederwall},
which can consist, in particular, of a flat background with
constant $B$ and $C$ fields. The case of vanishing $C$-field, for
which we know that a \nc\ description is possible, is mapped to
the generic one by \sz. Since \sz\ is a symmetry of IIB string
theory, should there not exist an \sz-covariant \nc\ description
of D3-branes in the presence of both constant $B$ and $C$ fields?

\vspace{1cm}
\noindent {\bf Note added:} While this paper was being written, I
learned about \cite{Hata}, which has some overlap with 
section \ref{nc-dyons-section}.

\vspace{1cm} 
\noindent {\bf Acknowledgments:} 
I am grateful to Joaquim Gomis for
the suggestion which motivated this work, and to both
him and Joan Sim\'on for illuminating discussions and useful
comments on the manuscript. I am also grateful to Selena Ng for
her helpful comments on a previous version of this paper.
Finally, I would like to thank Emili
Elizalde, Josep I. Latorre, Josep M. Pons, Toni Mateos and Jordi Molins for
discussions. 
This work has been supported by a
fellowship from the Comissionat per a Universitats i Recerca de la
Generalitat de Catalunya.
\newpage


\begin{thebibliography}{99}

\bibitem{Connes}
A. Connes, M. Douglas and A. Schwarz,
{\it Non-commutative Geometry and Matrix Theory: Compactification on
Tori}, JHEP {\bf 9802:003} (1998), hep-th/9711162

\bibitem{DH}
M. Douglas and C. Hull,
{\it D-branes and the Noncommutative Torus},
JHEP {\bf 9802:008} (1998), hep-th/9711165.

\bibitem{SW}
N. Seiberg and E. Witten,
{\it String Theory and Non-commutative Geometry},
JHEP {\bf 9909:032} (1999), hep-th/9908142.

\bibitem{CM}
C.G.~Callan and J.M.~Maldacena,
{\it Brane dynamics from the Born-Infeld Action},
\NP{B513} (1998) 198-212, hep-th/9708147.

\bibitem{Gibbons}
G.W. Gibbons, {\it Born-Infeld particles and Dirichlet p-branes},
\NP{B514} (1998) 603-639, hep-th/9709027.

\bibitem{BPSbounds}
J.P. Gauntlett, J. Gomis and P.K. Townsend,
{\it BPS bounds for Worldvolume Branes},
JHEP {\bf 9801:033} (1998), hep-th/9711205.

\bibitem{Thorlacius}
L. Thorlacius, 
{\it Born-Infeld String as a Boundary Conformal Field Theory},
\PRL{80} (1998) 1588-1590, hep-th/9710181.

\bibitem{Polchinski}
J. Polchinski, {\it Dirichlet-Branes and Ramond-Ramond Charges},
\PRL{75} (1995) 4724, hep-th/9510017.

\bibitem{Has-elec}
K. Hashimoto, {\it Born-Infeld Dynamics in Uniform Electric Field},
JHEP {\bf 9907:016} (1999), hep-th/9905162.

\bibitem{Has-Has}
A. Hashimoto and K. Hashimoto,
{\it Monopoles and Dyons in Non-commutative Geometry},
JHEP {\bf 9911:005} (1999), hep-th/9909202.

\bibitem{Brane-Config}
K. Hashimoto, H. Hata and S. Moriyama,
{\it Brane Configuration from Monopole Solution in Non-Commutative
Super Yang-Mills Theory}, 
JHEP {\bf 9912:021} (1999), hep-th/9910196.

\bibitem{Thorlacius.Peet}
S. Lee, A. Peet and L. Thorlacius, {\it Brane-Waves and Strings},
\NP{B514} (1998) 161-176, hep-th/9710097.

\bibitem{Asakawa}
T. Asakawa and I. Kishimoto,
{\it Comments on Gauge Equivalence in Noncommutative Geometry},
JHEP {\bf 9911:024} (1999), hep-th/9909139.

\bibitem{Dayi}
O. Dayi, {\it BRST-BFV analysis of equivalence between 
noncommutative and ordinary gauge theories},
hep-th/0001218.

\bibitem{Cederwall}
M. Cederwall and A. Westerberg,
{\it World-volume fields, \sz\ and duality: The type IIB 3-brane},
JHEP {\bf 9802:004} (1998), hep-th/9710007.

\bibitem{Hata}
H. Hata and S. Moriyama,
{\it String Junction from Non-Commutative Super Yang-Mills Theory},
hep-th/0001135.




\end{thebibliography}
\end{document}